\documentclass[paper]{ieice}

\usepackage[dvipdfmx]{graphicx}
\usepackage{subcaption}
\usepackage[fleqn]{amsmath}
\usepackage{newtxtext}
\usepackage[varg]{newtxmath}

\usepackage{mathtools,etoolbox}
\mathtoolsset{showonlyrefs,showmanualtags}

\newtheorem{theorem}{Theorem}
\newtheorem{lemma}{Lemma}

\field{}
\title{Sparse Superposition Codes with Binomial Dictionary are Capacity-Achieving with Maximum Likelihood Decoding}
\titlenote{This paper was presented in part at the 45th Symposium on Information Theory and its Applications (SITA2022).}
\authorlist{%
 \authorentry{Yoshinari TAKEISHI}{m}{labelA}\MembershipNumber{2203614}
 \authorentry{Jun'ichi TAKEUCHI}{m}{labelA}\MembershipNumber{}
}
\affiliate[labelA]{Faculty of Information Science and Electrical Engineering  Kyushu University,  Fukuoka 819-0395, Japan}

\begin{document}
\maketitle
\begin{summary}
It is known that sparse superposition codes asymptotically achieve the channel capacity over the additive white Gaussian noise channel with both maximum likelihood decoding and efficient decoding (Joseph and Barron in 2012, 2014). Takeishi et al. (in 2014, 2019) demonstrated that these codes can also asymptotically achieve the channel capacity with maximum likelihood decoding when the dictionary is drawn from a Bernoulli distribution. In this paper, we extend these results by showing that the dictionary distribution can be naturally generalized to the binomial distribution.
\end{summary}

\section{Introduction}
Sparse Superposition Codes (SS codes), proposed in 2010 \cite{Barron_f1,Barron_f2}, are error correction codes that form codewords through a sparse linear combination of the column vectors of a matrix called a dictionary. It was proven that they can asymptotically achieve the theoretical limit of transmission rate (channel capacity) in the additive white Gaussian noise (AWGN) channels with low decoding complexity \cite{Barron2}. In addition, when maximum likelihood decoding is used for SS codes, the channel capacity is achieved, and the decoding error probability decreases exponentially with the codelength. 
Although maximum likelihood decoding is computationally intractable, it is important to understand the theoretical performance limits of SS codes.

In SS codes, it is assumed that the dictionary used for encoding is generated independently from a certain Gaussian distribution for each element (referred to as a Gaussian dictionary). However, since the Gaussian distribution is unbounded, implementation in actual devices is difficult. There is also the challenge of determining how many bits should be used to quantize the continuous real values for storage in memory. In relation to this issue, Takeishi et al. proved that even when the dictionary is generated from a Bernoulli distribution (referred to as a Bernoulli dictionary), the channel capacity can still be achieved in maximum likelihood decoding \cite{Takeishi_P1,Takeishi1,Takeishi_P2,Takeishi2}. This is due to the distribution of the codewords approaching the same Gaussian distribution as the original, as the number of summations of the dictionary increases, according to the Central Limit Theorem. Consequently, this allows for storage of the dictionary in memory with just 1 bit per element, significantly simplifying the generation and management of the dictionary. However, it should be noted that the upper bound of the decoding error probability worsens slightly compared to the case of the Gaussian distribution, due to the error of the Central Limit Theorem.

In this paper, we demonstrate that even when the dictionary distribution is assumed to be binomial, the channel capacity can be achieved through maximum likelihood decoding, and the upper bound of the decoding error probability can be improved compared to the case of the Bernoulli distribution. The binomial distribution is a probability distribution that governs the number of successes in $d$ repeated Bernoulli trials, thus it can be considered a generalization of the Bernoulli distribution. By using a binomial distribution dictionary, the number of Bernoulli trials needed for a codeword is simply multiplied by $d$, making it relatively easy to generalize the proof for the case of Bernoulli distribution. Additionally, by increasing $d$, it is possible to reduce the error of the central limit theorem without changing the codelength. 
On the other hand, assuming fixed-length coding, the memory capacity required per dictionary element increases to $\log(d+1)$ bits.

As the number of trials increases, the binomial distribution approaches a Gaussian distribution, making it a {\it good} dictionary distribution among general discrete distributions in this sense. However, performance evaluation using these general discrete distributions remains a challenge. Takeishi et al. \cite{Takeishi_P2} noted that when using a Bernoulli distribution with unequal probabilities, which means the success probability of a trial differs from $1/2$, the proof could not proceed the same as with equal probabilities. The cause lies in the failure to uniformly evaluate the deference between the binomial distribution and the Gaussian distribution due to the distortion of the Bernoulli distribution. Therefore, to generalize to discrete distributions, it may be necessary to impose some assumptions regarding the symmetry of the distribution. Additionally, considering practical applications, the performance evaluation of discrete dictionaries for low-complexity decoding remains a task.

In the following sections, we first review the original SS codes, and then evaluate the upper bound on the error probability with the binomial dictionary. We also numerically verify the achievable rates for finite lengths according to the error bound. 

In this paper, a Gaussian distribution with mean $\mu$ and variance $\sigma^2$ is denoted as $N(\mu, \sigma^2)$. Moreover, 'log' refers to the logarithm with base 2, and 'ln' refers to the natural logarithm.

\section{Review of the original SS codes}
In this section, according to \cite{Barron1}, we review the performance of the original SS codes (with a Gaussian dictionary) through maximum likelihood decoding.

\subsection{Problem Setting}
We consider the problem of transmitting a $K$ bit message $u \in \{0,1\}^K$ over the AWGN channel. Each message occurs with equal probability $1/2^K$. The sender encodes $u$ to $c \in \mathbb{R}^n$, which serves as the $n$ input values to the channel. The transmission rate is defined as $R=K/n$ (bit/transmission).
Let the output of the channel be $Y \in \mathbb{R}^n$, then
\begin{align}
Y=c+\epsilon
\end{align}
holds, where the noise $\epsilon$ follows $N(0, \sigma^2)$ independently in each dimension. The average power of the input $c$ is defined as \[\frac{1}{n} \sum_{i=1}^n c_i^2,\] constrained to not exceed $P$. Under this constraint, the channel capacity is well-known \cite{Shannon} to be
\begin{align}
C=\frac{1}{2}\log(1+v)\ {\rm (bit/transmission)},
\end{align}
where $v=P/\sigma^2$ is the signal-to-noise ratio (SNR).

\subsection{Encoding}
We explain the encoding method of the SS code. First, the message $u$ is mapped to $\beta \in \{0,1\}^N$. The vector $\beta$ is divided into $L$ sections of size $M$, where exactly one element in each section is 1, and the remaining elements are 0. That is, considering a bijective mapping from $u$ to $\beta$,
\begin{align}
2^K=M^L \Leftrightarrow K=L\log M
\label{K2M}
\end{align}
holds. The codeword $c$ is then created as $c=X\beta$. Here, the matrix $X \in \mathbb{R}^{n \times N}$ is called a dictionary.

Each entry of the dictionary $X$ is generated independently from $N(0, P/L)$. 
Then, the distribution of each dimension of the codeword $c_i$ is independently follows $N(0, P)$, which maximizes the mutual information between the input and the output of the channel under the power constraint $P$.
Furthermore, to prevent the dictionary size $N$ from growing exponentially in $n$, we assume $M=L^a$, where $a$ is a constant called a section size rate. Then, from \eqref{K2M}, 
\[n=\frac{K}{R}=\frac{aL\log L}{R}\] 
holds, meaning that $n$ is approximately linear in $L$. Thus, the dictionary size $N=LM$ grows polynomially in $n$.

\subsection{Performance of Maximum Likelihood Decoding}
First, we explain the definition of maximum likelihood decoding as addressed in this paper. We define the set 
\[{\cal B}=\{\beta\in \{0,1\}^N : \beta {\rm\ has\ one\ 1\ in\ each\ section}\}.\] 
Then, maximum likelihood decoding is defined as
\begin{align}
\hat{\beta}=\arg \min_{\beta\in {\cal B}}\| Y-X\beta\|^2
\end{align}
where $\| \cdot \|$ denotes the Euclidean norm.

Let $\beta$ be transmitted over the channel.  Then, the block error in decoding is the event $\hat{\beta} \neq \beta$. Additionally, let 
\[mistakes \in \{0,1,\ldots,L\}\] 
denote the number of sections in $\hat{\beta}$ where the positions of the non-zero elements differ from those in $\beta$. The ratio $mistakes/L$ is called a section error rate. 
For $\alpha_0\in [0,1]$, we define the event 
\[{\cal E}_{\alpha_0}=\{mistakes \geq \alpha_0 L\}.\]
Here, we define $\mathbb{P}_\beta[\cdot]$ as a joint distribution of the noise $\epsilon$ and dictionary $X$ for given $\beta$.
Due to our assumption for the input distribution, any $\beta$ arises equal probability. Then, the average probability of the error conditioned on $X$ is given by 
\[\bar{\mathbb{P}}[{\cal E}_{\alpha_0}|X]=\frac{1}{M^L}\sum_{\beta\in {\cal B}}\mathbb{P}_\beta[{\cal E}_{\alpha_0}|X].\]
The derivation of the above probability is difficult, so we consider the expectation over the dictionary distribution as 
\begin{align}
\bar{\mathbb{P}}[{\cal E}_{\alpha_0}]=\mathbb{E}_X \bar{\mathbb{P}}[{\cal E}_{\alpha_0}|X]=\frac{1}{M^L}\sum_{\beta\in {\cal B}}\mathbb{P}_\beta[{\cal E}_{\alpha_0}].
\end{align}
By symmetry, $\mathbb{E}_X\mathbb{P}_\beta[{\cal E}_{\alpha_0}|X]$ are equal for all $\beta$. Then, for all $\beta\in {\cal B}$, we have 
\begin{align}
\bar{\mathbb{P}}[{\cal E}_{\alpha_0}]=\mathbb{P}_\beta[{\cal E}_{\alpha_0}] .
\end{align}

The following theorem evaluates the upper bound of the probability of the event ${\cal E}_{\alpha_0}$. To describe the theorem, we define
\begin{align}
w_v=\frac{v}{[4(1+v)^2]\sqrt{1+(1/4)v^3/(1+v)}}
\end{align}
and 
\[g(x)=\sqrt{1+4x^2}-1.\]
For any $x \geq 0$, 
$g(x) \geq \min\{\sqrt{2}x,x^2\}$
holds. Using these, we define
\begin{align}
h(\alpha,\Delta)=\min\left\{
\alpha w_v \Delta,\frac{1}{4}g\left(\frac{\Delta}{2\sqrt{v}}\right) \right\}.
\end{align}
Also, $a_{v,L}$ is a quantity that converges to a finite value as $L \rightarrow \infty$, with its detailed definition given in \cite{Barron1}.

\begin{theorem}[Joseph and Barron 2012]
\label{thm1}
Each element of the dictionary $X$ is independently generated from $N(0, P/L)$. If the section size rate $a \geq a_{v,L}$, then for any transmission rate $R<C$, we have 
\begin{align}
\bar{\mathbb{P}}[{\cal E}_{\alpha_0}]=e^{-nE(\alpha_0,R)},
\label{Ealpha}
\end{align}
where
\begin{align}
E(\alpha_0,R)\geq h(\alpha_0,C-R)-\frac{\ln (2L)}{n}.
\end{align}
\end{theorem}

We aim to make $\Delta_n = C - R$ approach 0 as $n$ increases, while achieving an exponentially small probability with respect to $n$ as given by \eqref{Ealpha}. To achieve this, since $h(\alpha,\Delta_n) = \Omega(\Delta_n^2)$, it suffices to satisfy $\Delta_n = \Omega((\log L)^{1/2}/n^{1/2})$. Additionally, by using composition with Reed-Solomon code \cite{RS} as an outer code, the block error probability can also be made exponentially small in $n$ with $\Delta_n = \Omega((\log L)^{1/2}/n^{1/2})$. For details, refer to \cite{Barron1}.

To prove this theorem, we use the following lemma, which evaluates the probability of the event 
\[E_l = \{mistakes = l\} \] 
for $l = 1, 2, \ldots, L$.
This lemma is used to evaluate
\begin{align}
{\mathbb{P}}_\beta[{\cal E}_{\alpha_0}]=\sum_{l \geq \alpha_0 L}{\mathbb{P}}_\beta[E_l].
\end{align}
To describe the lemma, we define the following. For $\alpha \in [0,1]$, define $C_\alpha = (1/2) \ln(1 + \alpha v)$. This matches $C$ when $\alpha = 1$. We also note that  $C_{\alpha}-\alpha C$ is a non-negative function with respect to $\alpha$, and it is 0 only when $\alpha = 0, 1$. For a positive quantity $\Delta$ and $\rho \in [-1, 1]$, define
\begin{align}
D(\Delta,1-\rho^2)=\max_{\lambda\geq 0}
\Bigl\{ \lambda\Delta + \frac{1}{2}\ln(1-\lambda^2(1-\rho^2)) \Bigr\}
\label{eq:D}
\end{align}
and
\begin{align}
D_1(\Delta,1-\rho^2)=\max_{0\leq\lambda\leq 1}
\Bigl\{ \lambda\Delta + \frac{1}{2}\ln(1-\lambda^2(1-\rho^2)) \Bigr\}.
\label{eq:D1}
\end{align}
Note that both of the above take non-negative values.

\begin{lemma}[Joseph and Barron 2012]
\label{lem1}
Each element of the dictionary $X$ is independently generated from $N(0, P/L)$. For given $l \leq L$, let $\alpha = l/L$. Then, we have 
\begin{align}
\mathbb{P}_\beta[E_l]\leq\min_{t_\alpha\in[0,C_{\alpha}-\alpha R]}err_{Gauss}(\alpha),
\end{align}
where 
\begin{align}
err_{Gauss}(\alpha)=&{_LC_{\alpha L}}
\exp\{ -nD_1(\Delta_{\alpha},1-\rho_1^2) \}\nonumber\\
+&\exp\{ -nD(t_{\alpha},1-\rho_2^2) \}\label{lemequ}
\end{align}
with $\Delta_{\alpha}=C_{\alpha}-\alpha R-t_{\alpha}$,
$1-\rho_1^2=\alpha(1-\alpha)v/(1+\alpha v)$, and
$1-\rho_2^2=\alpha^2 v/(1+\alpha^2 v)$.
\end{lemma}

\section{SS codes with binomial dictionary}
In this section, we generalize the evaluation of the SS codes with Bernoulli dictionary \cite{Takeishi1,Takeishi2}
 to the case of binomial dictionary.

\subsection{Method to generate dictionary}
For a natural number $d$, we prepare a matrix $\tilde{X} \in \mathbb{R}^{n \times dN}$, each element of which is generated from a Bernoulli distribution taking the value $\sqrt{P/(dL)}$ with probability 1/2 and $ -\sqrt{P/(dL)} $ with probability 1/2. Using this matrix $ \tilde{X} $, each element of the binomial dictionary $ X \in\mathbb{R}^{n \times N}$ is created by
\begin{align}
X_{ij} = \sum_{k=d(j-1)+1}^{dj}\tilde{X}_{ik}.
\label{gen_bindic}
\end{align}
As a result, each element $ X_{ij} $ independently follows a binomial distribution with $ d $ trials. The expected average power of the codeword is $ P $ as in the original case.

\subsection{Review of the Bernoulli dictionary case}
In the above method, the case of $ d=1 $ corresponds to the Bernoulli dictionary considered in \cite{Takeishi1,Takeishi2}. That is, each element of the dictionary is independently generated from a Bernoulli distribution taking the value $\sqrt{P/L}$ with probability 1/2 and $-\sqrt{P/L}$ with probability 1/2. In this case, as $ L $ becomes large, the distribution of the codeword approaches the Gaussian distribution as in the original case due to the central limit theorem. It was conjectured in \cite{Barron_f1,Barron1} that 
SS codes are capacity-achieving also in this case. This conjecture was positively resolved in \cite{Takeishi1,Takeishi2}. Furthermore, a Bernoulli dictionary version of Theorem \ref{thm1} was shown in their works, and it was demonstrated that the block error probability is exponentially small in $n$. However, the upper bound of the error probability shown in this evaluation is worse than the original upper bound due to the difference of dictionary distributions.

\subsection{Performance of Maximum Likelihood Decoding}
By using the binomial dictionary discussed in this paper, 
the upper bound of the error probability shown in Bernoulli dictionary case can be improved as \( d \) increases. This is demonstrated in the following theorem, which generalizes the Bernoulli dictionary version of Theorem \ref{thm1} shown in \cite{Takeishi2}.

\begin{theorem}
\label{thm2}
Each element of the dictionary \( X \) is independently generated from a binomial distribution with \( d \) trials as defined in \eqref{gen_bindic}. If the section size rate \( a \geq a_{v,L} \) and \( L \geq 2 \), then for any transmission rate \( R < C \), we have
\begin{align}
\bar{\mathbb{P}}[{\cal E}_{\alpha_0}] = e^{-nE(\alpha_0,R)},
\label{Ealpha2}
\end{align}
where
\begin{align}
E(\alpha_0,R) \geq h(\alpha_0, C - R) - \frac{\ln (2L)}{n} - \iota(L),
\end{align}
and
\(\iota(L) = \max\{\iota_1, \iota_2\}\). The quantities \(\iota_1\) and \(\iota_2\) are defined in Lemma \ref{lem2}.
\end{theorem}
 
The following lemma shows that \(\iota(L)=O(1/\sqrt{L'})\), where we define $L'=dL$. In other words, for the binomial dictionary case, we can see that an exponentially small block error probability can be achieved if \(\Delta_n =C-R= \Omega(1/(L')^{1/4})\) is satisfied.
Thus, as $L$ becomes large, $\Delta_n$ approaches to 0 faster than the Bernoulli dictionary case. 
\begin{lemma}
\label{lem2}
Each element of the dictionary \(X\) is independently generated from a binomial distribution with \(d\) trials as defined in \eqref{gen_bindic}. Let \(\alpha_0 \in (0,1]\) be a real number, and for a given \(l \leq L\), let \(\alpha = l/L\). Then, for any \(L \geq 2\) and \(l\) satisfying \(\alpha_0 \leq \alpha \leq 1\),
\begin{align}
\mathbb{P}_\beta[E_l] \leq \min_{t_\alpha \in [0, C_\alpha - \alpha R]} err_{Bin}(\alpha),
\end{align}
where 
\begin{align}
err_{Bin}(\alpha) = &{_LC_{\alpha L}}
\exp\{ -n(D_1(\Delta_\alpha, 1 - \rho_1^2) - \iota_1) \}\\
+&\exp\{ -n(D(t_\alpha, 1 - \rho_2^2) - \iota_2) \},
\end{align}
with \(\Delta_\alpha = C_\alpha - \alpha R - t_\alpha\),
\(1 - \rho_1^2 = \alpha(1 - \alpha)v/(1 + \alpha v)\), and
\(1 - \rho_2^2 = \alpha^2 v/(1 + \alpha^2 v)\). Also, \(\iota_1\) and \(\iota_2\) are defined as
\begin{align}
\iota_1 &= \ln ((1 + \iota_3)(1 + \max\{\iota_4, \iota_5\})) \\
\iota_2 &= \phi(dL) + \ln \left(1 + \frac{2\eta}{dL} \right)\label{iota2},
\end{align}
where
\begin{align}
1 + \iota_3 &= \max_{\alpha_0 L \leq l \leq L}
\left(e^{\phi(dl)}\left(1 + \frac{\eta (1 + v)}{dl}\right)\right),\label{iota3} \\
1 + \iota_4 &= \max_{\alpha_0 L \leq l \leq L - \sqrt{L/d}}
\left(e^{\phi(dl) + \phi(d(L - l))}\left(1 + \frac{\eta}{dl}\right)\right.\label{iota_4}\\
&\left.\ \ \ \ \ \ \ \ \ \ \ \ \ \ \ \ \  \ \ 
\times\left(1 + \frac{\eta}{d(L - l)}\right)\right), \\
1 + \iota_5 &= \max_{L - \sqrt{L/d} \leq l \leq L - 1}
\left(\frac{e^{\phi(dl)}}{\sqrt{1 - 1/\sqrt{dL}}}
\left(1 + \frac{\eta}{dl}\right)\right),\label{iota_5def}
\end{align}
and \(\eta = \sqrt{9/(8\pi e)}\). The function \(\phi\) is defined in Lemma \ref{lem3}.
\end{lemma}

The function \(\phi(L')\) is shown to be \(O(1/L')\) by Lemma \ref{lem3} presented later. Therefore, since \(\iota_1 = O(1/\sqrt{L'})\) and \(\iota_2 = O(1/L')\), \(\iota(L)\) in Theorem \ref{thm2} is \(O(1/\sqrt{L'})\).

To prove Lemma \ref{lem2}, similar to the Bernoulli dictionary case in \cite{Takeishi2}, the difference between the binomial distribution and the Gaussian distribution is evaluated using the following two lemmas.

\begin{lemma}[Takeishi et al. 2014]
\label{lem3}
For any natural number \(l\), we have
\begin{align}
\max_{k \in \{0,1,\ldots,l\}}\frac{{}_lC_k (1/2)^l}{N(k| l/2,l/4)}
\leq \exp\{\phi(l)\},
\label{lem3-1}
\end{align}
where
\begin{align}
N(x|\mu,\sigma^2) = \frac{1}{2\pi \sigma^2}\exp\left(-\frac{(x-\mu)^2}{2\sigma^2}\right),
\end{align}
\begin{align}
\phi(l) = \inf_{\zeta \in (0,1/2)} \phi_{\zeta}(l),
\end{align}
and
\begin{align}
\phi_{\zeta}(l) = \max\left\{ \left(\frac{3}{16}c_{\zeta}^2
+\frac{1}{12}\right)\frac{1}{l},\ 
 -\frac{4\zeta^4}{3}l + \ln\frac{l}{2} + \frac{1}{12l},\right.\\
 \left.
 -\left(\ln 2 - \frac{1}{2}\right)l + \frac{1}{2}\ln\frac{\pi l}{2}
 \right\},
\end{align}
where \(c_{\zeta} = 1/(1 + 2\zeta)^2 + 1/(1 - 2\zeta)^2\). 
Specifically, for any natural number \(l \geq 1000\),
\(\phi(l) \leq 5/l\)
holds.
\end{lemma}

\begin{lemma}[Takeishi and Takeuchi 2019]
\label{lem_sm}
For a natural number \(n\),
define \(h = 2/\sqrt{n}\) and
\begin{align}
{\cal X} = \{h(k - n/2) : k = 0,1,\ldots,n\}
\end{align}
Moreover, for \(s^2 > 0\) and \(\mu \in \Re\), define \(I_d\) and \(I_c\) as
\begin{align}
I_d = h \sum_{x \in {\cal X}} \exp\left\{ -\frac{s^2}{2}(x - \mu)^2 \right\}
\end{align}
\begin{align}
I_c = \int_{-\infty}^{\infty} \exp\left\{ -\frac{s^2}{2}(x - \mu)^2 \right\} dx = \sqrt{2\pi/s^2}
\end{align}
Then,
\begin{align}
I_d \leq \left(1 + \frac{\eta s^2}{n}\right)I_c
\end{align}
holds, where \(\eta = \sqrt{9/(8\pi e)} \leq 0.37\).
\end{lemma}

In \cite{Takeishi2}, Lemma \ref{lem_sm} is also shown for the cases where the variable \(x\) is two-dimensional and three-dimensional as follows.

\begin{lemma}[Takeishi and Takeuchi 2019]
\label{lem_sm2}
For a natural number $n$, define 
$h=2/\sqrt{n}$ and ${\cal X}=\{h(k-n/2)\mid k=0,1,\ldots,n\}$. 
Further, for a 2-dimensional real vector ${\bf x}=(x_1,x_2)^T$ 
and a strictly positive definite $2\times 2$ matrix $A$, define 
\[
I_d=\int_{-\infty}^{\infty}h\sum_{x_1\in {\cal X}}
\exp\left\{-\frac{{\bf x}^TA{\bf x}}{2}\right\}dx_2
\]
and 
\[
I_c=\int_{-\infty}^{\infty}\int_{-\infty}^{\infty}
\exp\left\{-\frac{{\bf x}^TA{\bf x}}{2}\right\}dx_1dx_2.
\]
Then, we have
\[
I_d\leq \left(1+\frac{\eta A_{11}}{n}\right)I_c, 
\]
where $\eta=\sqrt{9/(8\pi e)}\leq 0.37$ and $A_{11}$ is (1,1) element of 
matrix $A$.
\end{lemma}

\begin{lemma}[Takeishi and Takeuchi 2019]
\label{lem_sm3}
For natural numbers $n$ and $n'$, define 
${\cal X}_1=\{h_1(k-n/2)\mid k=0,1,\ldots,n\}$ and 
${\cal X}_2=\{h_2(k-n'/2)\mid k=0,1,\ldots,n'\}$, 
where $h_1=2/\sqrt{n}$ and $h_2=2/\sqrt{n'}$. 
Further, for a 3-dimensional real vector ${\bf x}=(x_1,x_2,x_3)^T$ 
and a strictly positive definite $3\times 3$ matrix $A$, define 
\[
I_d=\int_{-\infty}^{\infty}
h_1h_2\sum_{x_1\in {\cal X}_1}\sum_{x_2\in {\cal X}_2}
\exp\left\{-\frac{{\bf x}^TA{\bf x}}{2}\right\}dx_3
\]
and 
\[
I_c=\int_{\Re^3}
\exp\left\{-\frac{{\bf x}^TA{\bf x}}{2}\right\}d{\bf x}.
\]
Then, we have
\[
I_d\leq \left(1+\frac{\eta A_{11}}{n}\right)\left(1+\frac{\eta A_{22}}{n'}\right)I_c, 
\]
where $\eta=\sqrt{9/(8\pi e)}\leq 0.37$ and $A_{ij}$ is $(i,j)$ element of matrix $A$.
\end{lemma}

\subsection{Proof of Lemma \ref{lem2}}
In this proof, we replace the expectation of the binomial distribution dictionary \(X\) with the expectation of the Bernoulli dictionary \(\tilde{X}\), following the same steps as in the proof of Lemma 4 in \cite{Takeishi2}. The difference from the original proof is that the number of trials in the binomial distribution related to the dictionary is multiplied by \(d\). Note, however, that other parameters, such as the codelength \(n\) and the number of sections $L$, remain unchanged.

For $\beta\in \cal B$, let $S(\beta)=\{j: \beta_j=1\}$ be the index set of $\beta$.
Furthermore, define a set ${\cal A}=\{S(\beta): \beta\in {\cal B}\}$. Hereafter, let $\beta^*$ denote the $\beta$ sent to the channel, and let $S^*=S(\beta^*)$. 
Furthermore, let $X_S=\sum_{j\in S}X_j$, where $X_j$ denotes the $j$-th column vector of $X$. 

For the occurrence of the event $E_l$, there must be an $S\in {\cal A}$ which differs 
from $S^*$ in an amount $l$ and which has $\| Y-X_S\|^2
\leq \| Y-X_{S^*}\|^2$. 
Here, we define 
\begin{align}
T(S)=\frac{1}{2}\left[ \frac{| Y-X_{S} |^2}{\sigma^2}
-\frac{| Y-X_{S^*}| ^2}{\sigma^2} \right],
\end{align}
where for a vector $x \in \mathbb{R}^n$, 
$|x|^2$ denote $(1/n)\sum_{i=1}^n x_i^2$. 
Then, $\| Y-X_S\|^2\leq \| Y-X_{S^*}\|^2$ is equivalent to $T(S)\leq 0$. 
Here, we decompose $T(S)=\tilde{T}(S)+T^*$, where 
\begin{align}
\tilde{T}(S)=\frac{1}{2}
\left[ \frac{| Y-X_{S} |^2}{\sigma^2}
-\frac{| Y-(1-\alpha)X_{S^*}| ^2}{\sigma^2+\alpha^2P} \right]
\nonumber
\end{align}
and 
\begin{align}
T^*=\frac{1}{2}
\left[ \frac{| Y-(1-\alpha)X_{S^*} |^2}{\sigma^2+\alpha^2P}
-\frac{| Y-X_{S^*}| ^2}{\sigma^2} \right].
\nonumber
\end{align}
Introducing a parameter $t_\alpha \in [0, C_\alpha - \alpha R]$, define the event $\tilde{E}_l$ that there is an $S\in {\cal A}$ which differs 
from $S^*$ in an amount $l$ and $\tilde{T}(S)\leq t_\alpha$, and the event 
$E_l^*$ that $T^*\leq -t_\alpha=t^*$. Then, we have
\begin{align}
\mathbb{P}_{\beta^*}[E_l]=\mathbb{P}_{\beta^*}[E_l^*]+\mathbb{P}_{\beta^*}[\tilde{E}_l]
\end{align}

First, we evaluate $\mathbb{P}_{\beta^*}[E_l^*]$. From same discussion as the proof of Lemma \ref{lem1} (refer to \cite{Barron1}), for a parameter $\lambda$ such that 
$0 \leq \lambda<1/\sqrt{1-\rho_2^2}=1+1/\alpha^2v $, it follows that 
\begin{align}
\mathbb{P}_{\beta^*}[E_l^*]\leq e^{n\lambda t^*} \mathbb{E}_{\epsilon,X_{S^*}}e^{-n\lambda T^*}.
\nonumber
\end{align}
By using $\epsilon$ and $X_{S^*}$, we write down $T^*$ as 
\begin{align}
T^*=\frac{1}{2}
\left[ \frac{| \epsilon+\alpha X_{S^*} |^2}{\sigma^2+\alpha^2P}
-\frac{|\epsilon| ^2}{\sigma^2} \right].
\end{align}
Here, each coordinate of $X_{S^*}$ is subject to the binomial distribution with \(dL\) trials due to our way of making the dictionary. In the proof of Lemma 4 described in \cite{Takeishi2}, the number of trials for the corresponding binomial distribution is \(L\). 
Similarly to that proof, by writing down the expectation $\mathbb{E}_{\epsilon,X_{S^*}}e^{-n\lambda T^*}$ and applying Lemma \ref{lem3}, we have for ${\bf x}=(x_1,x_2)^T$
\begin{align}
\mathbb{P}_{\beta^*}[E_l^*]\leq e^{n\lambda t^*}
\left(\frac{e^{\phi(dL)}}{2\pi}
\int_{-\infty}^{\infty}h_1\sum_{x_1\in{\cal X}_1}
e^{-{\bf x}^T A {\bf x}/2}dx_2\right)^n, \label{dif_A}
\end{align}
where $h_1=2/\sqrt{dL}$, ${\cal X}_1=\{h_1(k-dL/2): k=0,1,\ldots,dL\}$, 
and $A=I-\lambda B$ with the identity matrix $I$ and
\begin{align}
B=(1-\rho_2^2)\begin{pmatrix} -1 & \frac{1}{\alpha\sqrt{v}} \\
 \frac{1}{\alpha\sqrt{v}} & 1 \end{pmatrix}.
\nonumber
\end{align}
Furthermore, by applying Lemma \ref{lem_sm2} to \eqref{dif_A}, which allows the replacement of the summation with the integral, we have
\begin{align*}
&\frac{e^{\phi(dL)}}{2\pi}
\int_{-\infty}^{\infty}h_1\sum_{x_1\in{\cal X}_1}
e^{-{\bf x}^T A {\bf x}/2}dx_2\\
\leq&
\frac{e^{\phi(dL)}}{2\pi}
\left(1+\frac{\eta A_{11}}{dL}\right)
\int_{-\infty}^{\infty}\int_{-\infty}^{\infty}
e^{-{\bf x}^T A {\bf x}/2}dx_1 dx_2\\
=&
\left(1+\frac{\eta A_{11}}{dL}\right)
\frac{e^{\phi(dL)}}{\sqrt{ 1-\lambda^2(1-\rho_2^2)}}\\
\leq & \frac{e^{\iota_2}}{\sqrt{ 1-\lambda^2(1-\rho_2^2)}},
\end{align*}
where we have used $A_{11}\leq 2$ and \eqref{iota2}.
Then, we have
\begin{align}
\!\!\!\!\mathbb{P}_{\beta^*}[E_l^*]&\leq e^{n\lambda t^*}
\left(\frac{e^{\iota_2}}{\sqrt{ 1-\lambda^2(1-\rho_2^2)}}\right)^n\nonumber\\
&=\exp\left\{-n\left(\lambda t_{\alpha}+\frac{1}{2}\ln(1-\lambda^2(1-\rho_2^2))-\iota_2\right)\right\}.\label{el_asta}
\end{align}

Second, we evaluate $\mathbb{P}_{\beta^*}[\tilde{E}_l]$.
The indicator of the event $\tilde{E}_l$
satisfies the inequality 
\[
1_{\tilde{E}_l} \leq \sum_{S_1}\Bigl(\sum_{S_2} e^{-n(\tilde{T}(S)-\tilde{t})} \Bigr)^\lambda,
\]
for an arbitrary $\lambda \in [0,1]$.
Here, $S_1=S\cap S^*$ of 
size $L-l$ and a difference $S_2=S\setminus S^*$
($=S\setminus S_1$) of size $l$. The outer sum in the above is over $S_1\subset S^*$, and for each $S_1$, the inner sum is among the $M-1$ choices in each $l$ sections determined by $S_1$.
By the same discussion
for $\mathbb{P}_{\beta^*}[E_l]$ in p.2547 of \cite{Barron1}, we have
\begin{align}
\!\!\mathbb{P}_{\beta^*}[\tilde{E}_l]\!\leq\!\sum_{S_1}\mathbb{E}_{\epsilon,X_{S^*}} 
e^{-n\lambda(\tilde{T}_1(S_1)-\tilde{t})} 
\!\Bigl(\sum_{S_2} \mathbb{E}_{X_{S_2}} e^{-n\tilde{T}_2(S)} \Bigr)
^{\lambda},\label{tilderE}
\end{align}
where we further decomposed $\tilde{T}(S)=\tilde{T}_1(S_1)+\tilde{T}_2(S)$ by using the following quantities;
\begin{align*}
\tilde{T}_1(S_1)&=\frac{1}{2}
\left[ \frac{| Y-X_{S_1} |^2}{\sigma^2+\alpha P}
-\frac{| Y-(1-\alpha)X_{S^*}| ^2}{\sigma^2+\alpha^2P} \right].
\end{align*}
\begin{align*}
\tilde{T}_2(S)&=\frac{1}{2}\left[ \frac{| Y-X_{S} |^2}{\sigma^2}
-\frac{| Y-X_{S_1}| ^2}{\sigma^2+\alpha P} \right].
\end{align*}
According to \cite{Barron1}, if $X_{ij}\sim N(0,P/L)$, we have
\[\mathbb{E}_{X_{S_2}} e^{-n\tilde{T}_2(S)}=e^{-nC_{\alpha}}.\]
On the other hand, in Bernoulli or binomial dictionary case, 
\begin{align}
\mathbb{E}_{X_{S_2}} e^{-n\tilde{T}_2(S)}=
 \frac{P_{Y| X_{S_1}}(Y| X_{S_1})}
{P^{(c)}_{Y| X_{S_1}}(Y| X_{S_1})}
 e^{-nC_{\alpha}},\label{proddc}
\end{align}
where $p^{(c)}_{Y| X_{S_1}}$ is the conditional probability density function
of $Y$ given $X_{S_1}$ in case $X_{ij}\sim N(0,P/L)$ \cite{Takeishi2}.
To evaluate the right-hand side of \eqref{proddc}, we will prove that \(P_{Y|X_{S_1}}(Y|X_{S_1})\) is nearly uniformly bounded by \(P^{(c)}_{Y|X_{S_1}}(Y| X_{S_1})\) for all \(Y\) and \(X_{S_1}\). Here, we define \(Y'=Y-X_{S_1}\) and define \(P_{Y'}\) as the probability density function for each coordinate of \(Y'\), and \(P_{Y'}^{(c)}\) as \(P_{Y'}\) in the case \(X_{ij} \sim N(0,P/L)\). Then, we have
\begin{align}
\frac{P_{Y| X_{S_1}}(Y| X_{S_1})}
{P_{Y| X_{S_1}}^{(c)}(Y| X_{S_1})}
=\prod_{i=1}^n \frac{P_{Y'}(Y'_i)}{P^{(c)}_{Y'}(Y'_i)}.\label{yprime}
\end{align}

Note that $Y'=X_{S^*\setminus S_1}+\epsilon$, where $X_{S^*\setminus S_1}$ is subject to a binomial distribution with $dl$ trials.
Here, define a set ${\cal X}_2=\{h_2(k-dl/2): k=0,1,\ldots,dl\}$
with $h_2=2/\sqrt{dl}$.
Then, by applying Lemma \ref{lem3}, we have
\begin{align*}
\!\!\!\!\!\!\!\!\!\!P_{Y'}(Y'_i)\leq 
\frac{e^{\phi(dl)}h_2}{2\pi\sqrt{\sigma^2}} 
\!\!\! \sum_{w_2\in {\cal X}_2}\!
\exp\left\{-\frac{a_2(w_2-a_3Y'_i)^2+a_4{Y'_i}^2}{2}\right\},
\end{align*}
where $a_2=1+\alpha v$, $a_3=\sqrt{\alpha v/\sigma^2}/a_2$ 
and $a_4=1/(\sigma^2a_2)$ $=(\sigma^2+\alpha P)^{-1}$.
Using Lemma \ref{lem_sm}, we have
\begin{align*}
&h_2
\! \sum_{w_2\in {\cal X}_2}\!
\exp\left\{-\frac{a_2(w_2-a_3Y'_i)^2+a_4{Y'_i}^2}{2}\right\}\\
\leq
& \left(1+\frac{\eta a_2}{dl}\right)
\int_{-\infty}^{\infty}
\exp\left\{-\frac{a_2(w_2-a_3Y'_i)^2+a_4{Y'_i}^2}{2}\right\}dw_2.
\end{align*}
Since $a_2\leq 1+v$, from \eqref{iota3}, we have 
\begin{align}
P_{Y'}(Y'_i)\leq (1+\iota_3) P_{Y'}^{(c)}(Y'_i).\label{dperc}
\end{align}
Then from (\ref{proddc}), we have
\begin{align}
\sum_{S_2}
\mathbb{E}_{X_{S_2}} e^{-n\tilde{T}_2(S)}&\leq 
\sum_{S_2}
(1+\iota_3)^n e^{-nC_{\alpha}}.\\
&\leq(1+\iota_3)^n e^{-n(C_{\alpha} - \alpha R)}.
\label{iota3neq}
\end{align}
The last inequality follows since the number of sum for $S_2$ is less than 
\[M^l = e^{l\ln M} = e^{\alpha L \ln M} 
= e^{\alpha nR}.
\]
From \eqref{tilderE}, we have
\begin{align}
\mathbb{P}_{\beta^*}[\tilde{E}_l]\leq (1+\iota_3)^n
\sum_{S_1}{\mathbb E}_{\epsilon,X_{S^*}}e^{-n\lambda\tilde{T}_1(S_1)}
e^{-n\lambda\Delta_{\alpha}},
\label{iota_3}
\end{align}
where $\Delta_{\alpha}=C_{\alpha}-\alpha R-t_{\alpha}$.

Now, we evaluate the right side of \eqref{iota_3}. 
We write down $\tilde{T}_1(S_1)$ as 
\begin{align}
\tilde{T}_1(S_1)=\frac{1}{2}
\left[ \frac{| \epsilon+ X_{S_3} |^2}{\sigma^2+\alpha P}
-\frac{|\epsilon+\alpha (X_{S_3}+X_{S_1})| ^2}{\sigma^2+\alpha^2 P} \right],
\end{align}
where $S_3=S^*\setminus S_1$.
Then, the expectation ${\mathbb E}_{\epsilon,X_{S^*}}e^{-n\lambda\tilde{T}_1(S_1)}$ is considered as an expectation of independent random variables $\epsilon$, $X_{S_3}$, and $X_{S_1}$.
Here, each coordinate of $X_{S_3}$ and $X_{S_1}$ are subject to the binomial distribution with $dl$ and $d(L-l)$, respectively.

Then, we will make case argument for (i) $l \leq L - \sqrt{L/d}$ and 
(ii) $l > L-\sqrt{L/d}$. 
Note that the condition of the case argument matches that in \cite{Takeishi2} when $d=1$.
First, we consider the case (i) $l \leq L - \sqrt{L/d}$.
In this case, letting $\tilde{l}=L-l$, $d\tilde{l}$ is larger than $\sqrt{dL}$. 
Similarly as \cite{Takeishi2}, by writing down the expectation ${\mathbb E}_{\epsilon,X_{S^*}}e^{-n\lambda\tilde{T}_1(S_1)}$ and applying Lemma \ref{lem3}, we have for 
 ${\bf x}=(x_1,x_2,x_3)^T$
\begin{align*}
& {\mathbb E}_{\epsilon, X_{S^*}}e^{-n\lambda\tilde{T}_1(S_1)}\\
&\leq 
\left(\frac{e^{\phi(dl)+\phi(d\tilde{l})}}{(2\pi)^{3/2}}
\int_{-\infty}^{\infty}\!\!h_2h_3\sum_{x_1\in{\cal X}_2}\!
\sum_{x_2\in{\cal X}_3}\!
e^{-{\bf x}^T \tilde{A} {\bf x}/2}dx_3\right)^n,
\end{align*}
where $h_3=2/\sqrt{d\tilde{l}}$, ${\cal X}_3=\{h_3(k'-d\tilde{l}/2): k'=0,1,\ldots,d\tilde{l}\}$, 
and $\tilde{A}=I-\lambda \tilde{B}$ with the identity matrix $I$ and
\begin{align}
\!\!\!\!\tilde{B}=
\left( 
\begin{array}{ccc}
\frac{\alpha v}{1+\alpha v}-
\frac{\alpha^3v}{1+\alpha^2 v}
&
-\frac{ \alpha^2\sqrt{\alpha(1-\alpha)}v}{1+\alpha^2 v}
&
\frac{\sqrt{\alpha v}}{1+\alpha v}
-\frac{\alpha\sqrt{\alpha v}}{1+\alpha^2 v}
\\
-\frac{ \alpha^2\sqrt{\alpha(1-\alpha)}v}{1+\alpha^2 v}&
\-\frac{\alpha^2(1-\alpha)v}{1+\alpha^2 v} &
-\frac{ \alpha\sqrt{(1-\alpha)v}}{1+\alpha^2 v}\\
\frac{\sqrt{\alpha v}}{1+\alpha v}
-\frac{\alpha\sqrt{\alpha v}}{1+\alpha^2 v}& 
-\frac{ \alpha\sqrt{(1-\alpha)v}}{1+\alpha^2 v}&
\frac{1}{1+\alpha v} -\frac{1}{1+\alpha^2 v}\\
\end{array} 
\right).
\nonumber
\end{align}
Applying Lemma \ref{lem_sm3}, we have

\begin{align*}
\!\!&\frac{e^{\phi(dl)+\phi(d\tilde{l})}}{(2\pi)^{3/2}}
\int_{-\infty}^{\infty}\!\!h_2h_3\sum_{x_1\in{\cal X}_2}\!
\sum_{x_2\in{\cal X}_3}\!
e^{-{\bf x}^T \tilde{A} {\bf x}/2}dx_3\\
&\leq \frac{e^{\phi(dl)+\phi(d\tilde{l})}}{(2\pi)^{3/2}}
\left(1+\frac{\eta \tilde{A}_{11}}{dl}\right)\left(1+\frac{\eta \tilde{A}_{22}}{d\tilde{l}}\right)
\int_{\mathbb{R}^3}
e^{-{\bf x}^T \tilde{A} {\bf x}/2}d{\bf x}\\
&\leq \left(1+\frac{\eta \tilde{A}_{11}}{dl}\right)\left(1+\frac{\eta \tilde{A}_{22}}{d\tilde{l}}\right)
\frac{e^{\phi(dl)+\phi(d\tilde{l})}}
{\sqrt{ 1-\lambda^2(1-\rho_1^2)}}\\
&\leq \frac{1+\iota_4}
{\sqrt{ 1-\lambda^2(1-\rho_1^2)}},
\end{align*}
where we used $\tilde{A}_{11}\leq 1$, $\tilde{A}_{22}\leq 1$, and \eqref{iota_4}.
Thus, we have
\begin{align}
{\mathbb E}_{\epsilon,X_{S^*}}e^{-n\lambda\tilde{T}_1(S_1)}
\leq \left(\frac{1+\iota_4}
{\sqrt{ 1-\lambda^2(1-\rho_1^2)}}\right)^n.\label{iota4}
\end{align}

Next, we consider the case (ii) $l > L-\sqrt{L/d}$.
Since $d\tilde{l}$ can be small in this case, we use an alternative method for evaluating the expectation ${\mathbb E}_{\epsilon,X_{S^*}}e^{-n\lambda\tilde{T}_1(S_1)}$ as in the corresponding part of \cite{Takeishi2}.
Then, recalling  \eqref{iota_5def}, we have
\begin{align}
{\mathbb E}_{\epsilon,X_{S^*}}e^{-n\lambda\tilde{T}_1(S_1)}
\leq \left(\frac{1+\iota_5}
{\sqrt{ 1-\lambda^2(1-\rho_1^2)}}\right)^n.\label{iota5}
\end{align}

Then from \eqref{iota_3}, \eqref{iota4} and \eqref{iota5}, we have
\begin{align*}
\mathbb{P}_{\beta^*}[\tilde{E}_l]
&\leq 
(1+\iota_3)^n{}_LC_{\alpha L}\frac{(1+\max(\iota_4,\iota_5))^n}
{ (1-\lambda^2(1-\rho_1^2))^{n/2}}e^{-n\lambda\Delta_{\alpha}}\\
&={}_LC_{\alpha L}
e^{-n(\lambda\Delta_{\alpha}+(1/2)\ln(1-\lambda^2(1-\rho_1^2))-\iota_1)},
\end{align*}
where $\iota_1=\ln((1+\iota_3)(1+\max(\iota_4,\iota_5))$. 
Minimizing the right side for $0\leq \lambda\leq 1$, we have 
\begin{align}
\mathbb{P}_{\beta^*}[\tilde{E}_l]\leq
{}_LC_{\alpha L}
\exp\{-n(D_1(\Delta_{\alpha},1-\rho_1^2)- \iota_1)\}.
\label{el_tilde}
\end{align}
Therefore, the lemma is proven from \eqref{el_asta} and \eqref{el_tilde}.

\section{Numerical Calculations}
In the original SS codes, the maximum transmission rate that can achieve a block error probability of $\epsilon_0 = 10^{-4}$ or less with a finite codelength $n$ has been confirmed through numerical calculations \cite{Barron1}. The results of performing this calculation for the binomial dictionary discussed in this paper are shown in Fig. \ref{calc}. From these results, it can be seen that for both $v=20$ and $v=100$, when $d=1000$, achievable rates very close to those of the original case can be obtained. However, in the case of the Bernoulli dictionary ($d=1$), no useful upper bound on the error probability was obtained.
The details of the calculation method are described in the following subsection.

\subsection{Details of the Calculation}
\begin{figure}[htbp]
    \begin{tabular}{c}
      \begin{minipage}[t]{0.47\hsize}
        \centering
        \includegraphics[width=8.5cm]{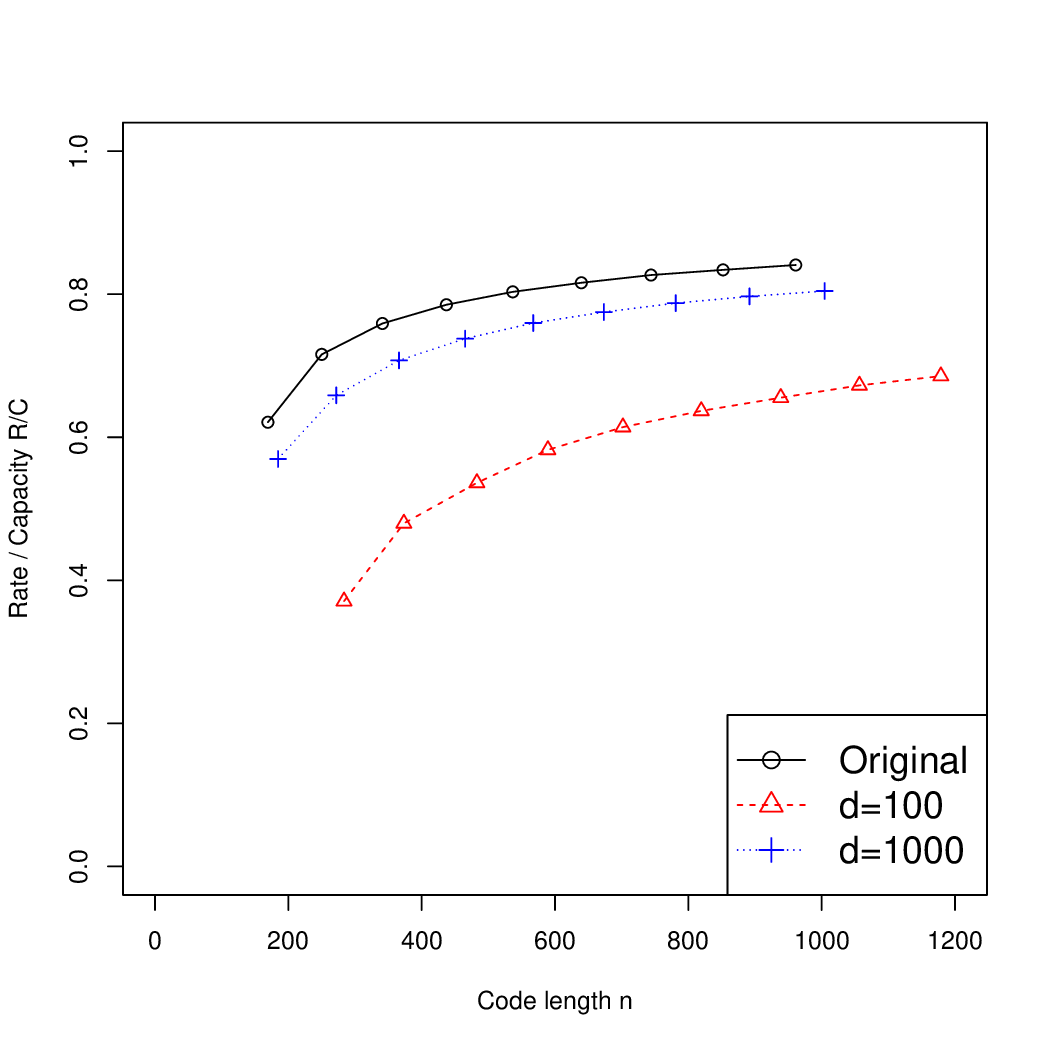}
        \subcaption{When SNR is $v=20$}
      \end{minipage} \\
      \begin{minipage}[t]{0.47\hsize}
        \centering
        \includegraphics[width=8.5cm]{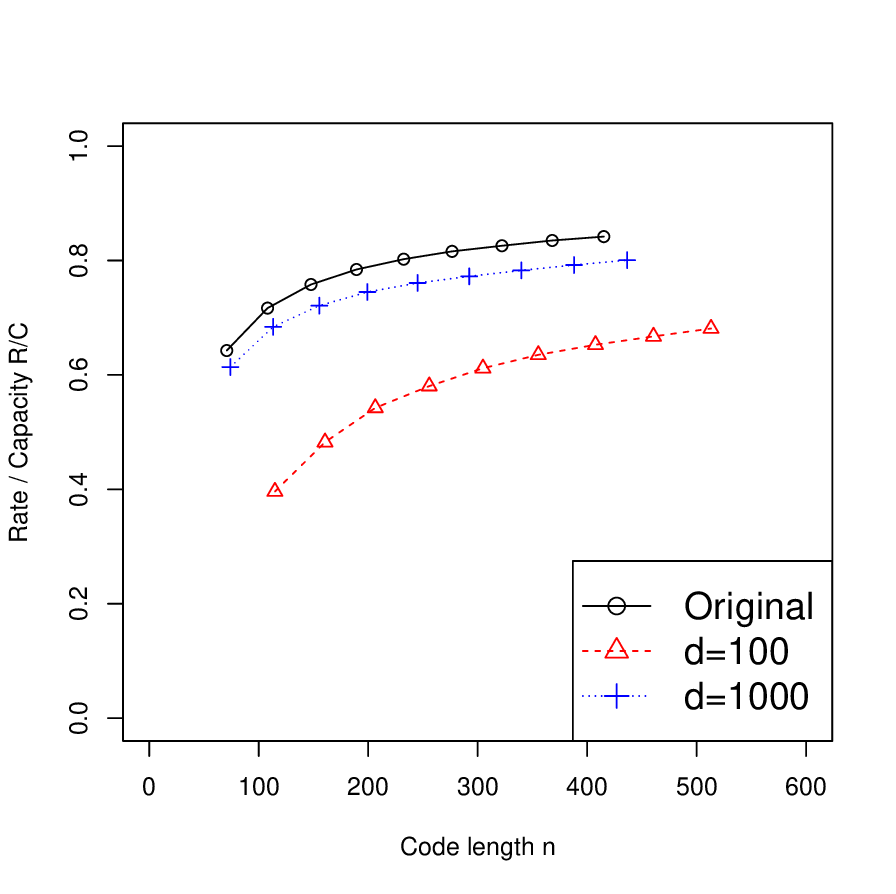}
        \subcaption{When SNR is $v=100$}
      \end{minipage} 
    \end{tabular}
    \caption{Graphs of achievable rates for the original normal distribution dictionary and the binomial distribution dictionary ($d=100$, $d=1000$).
    }
    \label{calc}
  \end{figure}

In Fig. \ref{calc}, $L$ was varied from 20 to 100 in increments of 10, and for each $L$, we calculated the maximum rate that could achieve a block error probability of $\epsilon_0 = 10^{-4}$ or less. 
The details of the calculation are described below. For each \( L \), \( R = K/n \) is varied from \( 0.99C \) to \( 0.05C \) in decrements of \( 0.001C \), and for each \( R \), the minimum \( l \) that satisfies \( \bar{\mathbb{P}}[{\cal E}_{\alpha_0}] \leq \epsilon_0 \) is determined. By combining an outer code with a rate of \( 1-2\alpha_0 \) at this point, a block error probability of \( \epsilon_0 \) can be achieved. Among the tested values of \( R \), the maximum overall rate \( (1-2\alpha_0)R \) is considered the achievable rate for each \( L \).
We plotted the corresponding codelength $n$ and achievable rate for each $L$.
This calculation follows the method described in APPENDIX D of \cite{Barron1}.

The upper bound on the error probability used in the calculations in Figure \ref{calc} is the smaller of the upper bounds from Lemma \ref{lem2} and the following lemma:

\begin{lemma}
\label{lem2-1}
Each element of the dictionary \(X\) is independently generated from a binomial distribution with \(d\) trials as defined in \eqref{gen_bindic}.
 Let $\alpha_0 \in (0,1]$ be a real number, and for a given $l \leq L$, let $\alpha = l/L$. Then, for any $l$ satisfying $\alpha_0 \leq \alpha \leq 1$, it holds that $\Pr[E_l] \leq err_{Bin'}(\alpha)$, where
\begin{align}
err_{Bin'}(\alpha) = &{_LC_{\alpha L}}
\exp\{ -n(D_1(\Delta_{\alpha},1-\rho_\alpha^2)-\iota') \}
\end{align}
and $\Delta_{\alpha} = C_{\alpha} - \alpha R$,
$1-\rho_\alpha^2 = \alpha v/(1+\alpha v)$.
Moreover, $\iota'$ is defined by
\begin{align}
\iota' = \max_{\alpha_0 L\leq l \leq L}
\Bigl\{2\phi(dl)+\left.\ln\left(\left(1+\frac{\eta(1+v)}{dl} \right)\left(1+\frac{2\eta}{dl} \right)\right)\right\},
\end{align}
where $\eta = \sqrt{9/(8\pi e)}$, and the function $\phi$ is defined in Lemma $\ref{lem3}$.
\end{lemma}

In this lemma, $\iota' = O(1/L)$, and excluding this term corresponds to Lemma 3 in the original case \cite{Barron1}. These lemmas are not used in the proofs of Theorem \ref{thm1} or Theorem \ref{thm2}, but they are useful to evaluate the probability of the event where the section error rate $\alpha$ is not close to 1.

\subsection{Proof of Lemma \ref{lem2-1}}
The proof of Lemma \ref{lem2-1} is conducted in the same way as the proof of Lemma \ref{lem2}, by evaluating the error between the binomial and normal distributions to reduce the proof to the original case. 
Similarly to \eqref{tilderE}, we have for arbitrary $\lambda\in[0,1]$
\begin{align}
\!\!\mathbb{P}_{\beta^*}[{E}_l]\!\leq\!\sum_{S_1}\mathbb{E}_{\epsilon,X_{S^*}} 
e^{-n\lambda{T}_1(S_1)} 
\!\Bigl(\sum_{S_2} \mathbb{E}_{X_{S_2}} e^{-n{T}_2(S)} \Bigr)
^{\lambda},\label{tilderE-2}
\end{align}
where in this proof, we decomposed ${T}(S)={T}_1(S_1)+{T}_2(S)$ by using the following quantities;
\begin{align*}
{T}_1(S_1)&=\frac{1}{2}
\left[ \frac{| Y-X_{S_1} |^2}{\sigma^2+\alpha P}
-\frac{| Y-X_{S^*}| ^2}{\sigma^2} \right].
\end{align*}
\begin{align*}
{T}_2(S)&=\frac{1}{2}\left[ \frac{| Y-X_{S} |^2}{\sigma^2}
-\frac{| Y-X_{S_1}| ^2}{\sigma^2+\alpha P} \right]=\tilde{T}_2(S).
\end{align*}
By using \eqref{iota3neq}, we have
\begin{align}
\!\!\mathbb{P}_{\beta^*}[{E}_l]\!\leq\!(1+\iota_3)^n\sum_{S_1}\mathbb{E}_{\epsilon,X_{S^*}} 
e^{-n\lambda{T}_1(S_1)} 
 e^{-n(C_{\alpha} - \alpha R)}.
\end{align}
By writing down the expectation ${\mathbb E}_{\epsilon,X_{S^*}}e^{-n\lambda{T}_1(S_1)}$ and applying Lemma \ref{lem3}, we have for 
 ${\bf x}=(x_1,x_2)^T$
\begin{align*}
{\mathbb E}_{\epsilon, X_{S^*}}e^{-n\lambda\tilde{T}_1(S_1)}\leq 
\left(\frac{e^{\phi(dl)}}{2\pi}
\int_{-\infty}^{\infty}\!\!h_2\sum_{x_1\in{\cal X}_2}\!
e^{-{\bf x}^T {A'} {\bf x}/2}dx_2\right)^n,
\end{align*}
where $h_2=2/\sqrt{dl}$, ${\cal X}_2=\{h_2(k-dl/2): k=0,1,\ldots,dl\}$, 
and ${A'}=I-\lambda {B'}$ with the identity matrix $I$ and
\begin{align}
B'=(1-\rho_\alpha^2)\begin{pmatrix} -1 & \frac{-1}{\sqrt{\alpha v}} \\
 \frac{-1}{\sqrt{\alpha v}} & 1 \end{pmatrix}.
\nonumber
\end{align}
Furthermore, by applying Lemma \ref{lem_sm2} to the above inequality, we have
\begin{align*}
&\frac{e^{\phi(dl)}}{2\pi}
\int_{-\infty}^{\infty}\!\!h_2\sum_{x_1\in{\cal X}_2}\!
e^{-{\bf x}^T {A'} {\bf x}/2}dx_2\\
\leq&
\frac{e^{\phi(dl)}}{2\pi}
\left(1+\frac{\eta A'_{11}}{dl}\right)
\int_{-\infty}^{\infty}\int_{-\infty}^{\infty}
e^{-{\bf x}^T A' {\bf x}/2}dx_1 dx_2\\
=&
\left(1+\frac{2\eta}{dl}\right)
\frac{e^{\phi(dl)}}{\sqrt{ 1-\lambda^2(1-\rho_\alpha^2)}}
\end{align*}
where we have used $A'_{11}=1+\lambda(1-\rho_\alpha^2)\leq 2$. Then, we have
\begin{align*}
\mathbb{P}_{\beta^*}[{E}_l]
&\leq 
(1+\iota_3)^n {}_LC_{\alpha L}\frac{ \left(1+\frac{2\eta}{dl}\right)^ne^{n\phi(dl)}}
{ (1-\lambda^2(1-\rho_\alpha^2))^{n/2}}e^{-n\lambda\Delta_{\alpha}}\\
&={}_LC_{\alpha L}
e^{-n(\lambda\Delta_{\alpha}+(1/2)\ln(1-\lambda^2(1-\rho_\alpha^2))-\iota')}.
\end{align*}
Minimizing the right side for $0\leq \lambda\leq 1$, we derive the lemma.

\section{Conclusion}
We proved that the error probability of sparse superposition codes is exponentially small with binominal dictionary at rates up to capacity.
By increasing the number of trials of the binominal distribution, the derived error bound becomes closer to that with Gaussian dictionary.

\section*{Acknowledgment}
This work was supported by JSPS KAKENHI Grant Number JP24K17278.


\begin{thebibliography}{99}

\bibitem{Barron_f1}
A. R. Barron and A. Joseph, ``Least squares superposition coding of 
moderate dictionary size, reliable at rates up to channel capacity,'' 
{\it Proc. Int. Symp. Inf. Theory}, Austin, Texas, June 13--18, 2010, 
pp. 275--279.

\bibitem{Barron_f2}
A. R. Barron and A. Joseph, ``Towards fast reliable communication at rates 
near capacity with Gaussian noise,'' {\it Proc. IEEE. Int. Symp. Inf. 
Theory}, Austin, Texas, Jun. 13--18, 2010, pp. 315--319.

\bibitem{Barron1}
A.\ Joseph and A.\ R.\ Barron, ``Least squares superposition codes of 
moderate dictionary size are reliable at rates up to capacity,'' 
{\it IEEE Trans. Inf. Theory}, vol. 58, no. 5, pp. 2541--2557, May 2012.

\bibitem{Barron2}
A.\ Joseph and A.\ R.\ Barron, ``Fast sparse superposition codes have 
near exponential error probability for $R<{\cal C}$. 
{\it IEEE Trans. Inf. Theory}, vol. 60, no. 2, pp. 919--942, Feb 2014.

\bibitem{RS}
I. S. Reed and G. Solomon, ``Polynomial codes over certain finite
fields,'' {\it J. SIAM}, vol. 8, pp. 300--304, Jun. 1960.

\bibitem{Shannon}
C. E. Shannon,
``A mathematical theory of communication,''
{\it Bell Syst. Tech. J.,} vol.27, pp. 379--423, 1948.

\bibitem{Takeishi_P1}
Y. Takeishi, M. Kawakita, and J. Takeuchi,
``Least Squares Superposition Codes with Bernoulli Dictionary are Still Reliable at Rates up to Capacity,'' 
{\it Proc. of 2013 IEEE International Symposium on Information Theory}, 
pp. 1396--1400, Istanbul, Turkey, July 7-12, 2013.

\bibitem{Takeishi1}
Y. Takeishi, M. Kawakita, and J. Takeuchi, 
``Least squares superposition codes with Bernoulli dictionary are still 
reliable at rates up to capacity,'' 
{\it IEEE Trans. Inf. Theory}, vol. 60, no. 5, pp. 2737--2750, May 2014.

\bibitem{Takeishi_P2}
Y. Takeishi and J. Takeuchi,
``An Improved Upper Bound on Block Error Probability of Least Squares Superposition Codes with 
Unbiased Bernoulli Dictionary,'' {\it Proc. of 2016 IEEE International Symposium on Information Theory}, 
pp. 1168--1172, Barcelona, Spain, July 10-15, 2016.

\bibitem{Takeishi2}
Y. Takeishi and J. Takeuchi, ``An Improved Analysis of Least Squares Superposition Codes with Bernoulli Dictionary,'' 
{\it Japanese Journal of Statistics and Data Science}, 2, pp. 591--613, Sep. 2019.

\end{thebibliography}


\end{document}